\documentstyle[12pt]{article}              
\begin{document}
\title{Collectivity of the low-lying dipole strength in relativistic
random phase approximation}
\author{ D. Vretenar$^{1,2}$, N. Paar$^{1}$, 
P. Ring$^{1}$, and G.A. Lalazissis$^{1,3}$
\vspace{0.5 cm}\\
$^{1}$ Physik-Department der Technischen Universit\"at M\"unchen,\\
D-85748 Garching, Germany\\
$^{2}$ Physics Department, Faculty of Science, University of
Zagreb,\\ 10000 Zagreb, Croatia\\
$^{3}$Physics Department, Aristotle University of Thessaloniki,\\ 
Thessaloniki GR-54006, Greece\\
}
\bigskip
\maketitle
\begin{abstract}
The relativistic random phase approximation is applied in the analysis
of the evolution of the isovector dipole response in nuclei with a 
large neutron excess. The self-consistent framework of relativistic 
mean-field theory, which has been very successfully applied in the 
description of ground-state properties of nuclei far from the valley of 
$\beta$-stability, is extended to study the possible onset of 
low-energy collective isovector dipole modes in nuclei with extreme 
isospin values. 
\end{abstract}
\vspace{2cm}
\vspace{2cm}
\vspace{1 cm}
\newpage
\baselineskip = 24pt
\section{Introduction}

Exotic nuclei with a large neutron excess exhibit unique structure 
phenomena: the weak binding of the outermost
neutrons, pronounced effects of the coupling between bound
states and the particle continuum, regions of nuclei
with very diffuse neutron densities, the formation of neutron
skin and halo structures. The modification of the
effective nuclear potential results in a
suppression of shell effects, the disappearance
of spherical magic numbers, and the onset of deformation
and shape coexistence. Isovector quadrupole
deformations are expected at the neutron drip-lines, and possible
low-energy collective isovector modes have been predicted.

The multipole response of nuclei with large neutron excess has
been the subject of many theoretical studies in recent years.
In particular, studies of low-energy collective isovector modes
provide important information on the isospin and density dependent
parts of the effective interactions used in nuclear structure 
models. Namely, the isoscalar terms of the effective forces
are rather well determined by the ground-state properties 
(sizes, shapes and binding energies). The isovector channel,
on the other hand, is much less explored. At
present there are many more theoretical predictions than 
experimental data, and moreover, almost all available data are on 
light nuclei. For example, low-lying dipole strength was observed for 
the halo nuclei $^{6}$He~\cite{Aum.99}, $^{11}$Li~\cite{Zin.97}, 
and $^{11}$Be~\cite{Nak.94}. For $^{6}$He and $^{11}$Li about 10\% of 
the Thomas-Reiche-Kuhn (TRK) dipole sum rule was found below 
an excitation energy of 5 MeV. For $^{6}$He experimental evidence 
was also reported for low-lying multipole strength other than dipole.
Low-lying dipole strength, exhausting about 5\% of the TRK sum rule
for energies up to 5 MeV above the threshold, was observed 
for $^{18}$O, $^{20}$O and  $^{22}$O~\cite{Aum.99}.
The experimental results indicate that, for nuclei with a
large neutron excess, the strength distribution consists of 
two major parts. One part corresponds to the real collective 
excitation, the other at lower excitation energy is caused by
single particle excitations of loosely bound neutrons. These
results, however, do not exclude the onset of collectivity 
in the low-energy region for the multipole response of 
medium-heavy and heavy nuclei.

Most theoretical studies of the multipole response of exotic 
nuclei have addressed the following problem: What happens to
giant resonances in nuclei with a large neutron excess?
Catara {\it et al.}
have studied the low-lying components in strength distributions
of weakly bound neutron-rich nuclei~\cite{Cat.96}, and the effect of
large neutron excess on the dipole response in the region of the giant
dipole resonance in O and Ca isotopes~\cite{Cat.97}. They have shown that
the neutron excess increases the fragmentation of the isovector giant
dipole resonance (GDR) and that the radial separation of proton and neutron
densities leads to non-vanishing isoscalar transition densities to
the GDR states. The fragmentation of the
isoscalar and isovector monopole strength
in neutron rich Ca isotopes has been studied in Ref.~\cite{HSZ.97}, and
in Ref.~\cite{Cha.94} the onset of soft dipole resonances in Ca isotopes
has been analyzed. Quite generally, the theoretical analyses have shown 
that the spectral distributions of exotic nuclei are much more 
fragmented than those for well bound systems. This happens because
protons and neutrons in exotic nuclei feel very different effective 
potentials, and consequently display very different level spacings.
This effect hinders the coherent accumulation of all strength in 
one dominant resonance mode~\cite{Rei.99}. In all calculations 
additional strength has been found below the normal giant resonance
region. The detailed predictions for the strength function, however, 
strongly depend on the effective forces used in the calculations. 
At this stage, therefore, it is important to compare results
of various theoretical analyses of the multipole response of 
exotic nuclei with a large ratio of neutron to proton numbers.

In the present study we apply the relativistic random phase 
approximation (RRPA) in the analysis of the isovector dipole 
response of nuclei with a large neutron excess. 
The self-consistent framework of relativistic 
mean-field theory, which has been very successfully applied in the 
description of ground-state properties of nuclei far from the valley of 
$\beta$-stability, is extended to study the possible onset of 
low-energy collective isovector modes in nuclei with extreme isospin
values. 

The low-energy collective isovector dipole mode, i.e. the pygmy
resonance, results from the excess neutrons oscillating out of phase
with a core composed of equal number of protons and neutrons.
A number of theoretical models have been applied in studies of the
dynamics of pygmy dipole resonances. These include: the three-fluid
hydrodynamical model (the protons, the neutrons of the same orbitals as
protons, and the excess neutrons)~\cite{MDB.71}, the two-fluid (the core
fluid and the neutron excess fluid) Steinwedel-Jensen hydrodynamical
model~\cite{Suz.90}, density functional theory~\cite{Cha.94}, and
the Hartree-Fock plus random phase approximation (RPA) with Skyrme
forces~\cite{Cat.97,ACS.96}.
More recently, large scale shell model calculations have been
performed in studies of pygmy and dipole states in O
isotopes~\cite{SS.99}, and dipole and spin-dipole strength distributions
in $^{11}$Li~\cite{SSB.00}.

There is also some experimental evidence of possible pygmy dipole states
in $^{208}$Pb. Studies of the low energy spectrum by elastic photon
scattering~\cite{SAC.82}, photoneutron~\cite{BCA.82}, and electron
scattering~\cite{Kuh.81} have detected fragmented E1 strength in the
energy region between 9 and 11 MeV. The fine structure exhausts
between 3 and 6\% of the E1 sum rule.
In a recent study~\cite{Pyg.00} of the isovector dipole response 
in $^{208}$Pb in the RRPA framework, we have found low-lying
E1 peaks in the energy region between 7 and 11 MeV. In particular,
a collective state has been identified whose dynamics correspond
to that of a dipole pygmy resonance. 

In this work we extend the analysis of Ref.~\cite{Pyg.00} and study the
evolution of collectivity in the isovector dipole response
in the low-energy region below the giant dipole resonance. The analysis
includes neutron rich isotopes of O, Ca, Ni, Zr, and Sn. In Section 2
we present an outline of the relativistic mean-field model and the 
relativistic random phase approximation. In Section 3 the model is 
applied in the analysis of the low-lying isovector dipole strength in 
neutron rich O and Ca isotopes. The onset of collective low-energy 
dipole states, i.e. pygmy resonances, in medium-heavy nuclei is
illustrated in Section 4 for Ni and Sn isotopes. The results are
summarized in Section 5.
\section{Isovector dipole excitations in the relativistic RPA}

Models based on quantum hadrodynamics provide a framework
in which the nuclear system is described by interacting nucleons and mesons.
In comparison with conventional non relativistic descriptions,
relativistic models explicitly include mesonic
degrees of freedom and consider the nucleons as Dirac particles.
A variety of nuclear phenomena have been described in the
relativistic framework: nuclear matter, properties of finite
spherical and deformed nuclei, hypernuclei, neutron stars,
nucleon-nucleus and electron-nucleus
scattering, relativistic heavy-ion collisions.
In particular, relativistic models
based on the mean-field approximation have
been successfully applied in the description of properties
of spherical and deformed $\beta$-stable nuclei,
and more recently in studies of exotic nuclei far from the
valley of beta stability.

The relativistic mean field theory is based on simple concepts:
nucleons are described as point particles, the theory is fully Lorentz
invariant, the nucleons move independently in
mean fields which originate from the nucleon-nucleon interaction.
Conditions of causality and Lorentz invariance impose that the
interaction is mediated by the
exchange of point-like effective mesons, which couple to the nucleons
at local vertices. The single-nucleon dynamics is described by the
Dirac equation
\begin{equation}
\label{statDirac}
\left\{-i\mbox{\boldmath $\alpha$}
\cdot\mbox{\boldmath $\nabla$}
+\beta(m+g_\sigma \sigma)
+g_\omega \omega^0+g_\rho\tau_3\rho^0_3
+e\frac{(1-\tau_3)}{2} A^0\right\}\psi_i=
\varepsilon_i\psi_i.
\end{equation}
$\sigma$, $\omega$, and
$\rho$ are the meson fields, and $A$ denotes the electromagnetic potential.
$g_\sigma$ $g_\omega$, and $g_\rho$ are the corresponding coupling
constants for the mesons to the nucleon.
The lowest order of the quantum field theory is the {\it
mean-field} approximation: the meson field operators are
replaced by their expectation values. The sources
of the meson fields are defined by the nucleon densities
and currents.  The ground state of a nucleus is described
by the stationary self-consistent solution of the coupled
system of the Dirac~(\ref{statDirac}) and Klein-Gordon equations.
The source terms in the Klein-Gordon equations for the meson fields
are calculated in the {\it no-sea} approximation, i.e. the Dirac sea
of negative energy states does not contribute to the nucleon densities
and currents. 
The quartic potential
\begin{equation}
U(\sigma )~=~\frac{1}{2}m_{\sigma }^{2}\sigma ^{2}+\frac{1}{3}g_{2}\sigma
^{3}+\frac{1}{4}g_{3}\sigma ^{4}  \label{usigma}
\end{equation}
introduces an effective density dependence. The non-linear
self-interaction of the $\sigma$ field is essential for
a quantitative description of properties of finite nuclei.

By adjusting just a few model parameters:
coupling constants and effective masses, to global properties of simple
spherical nuclei, it has been possible to describe many nuclear structure
phenomena, not only in nuclei along the valley of
$\beta$-stability, but also in exotic nuclei with extreme isospin
values and close to the particle drip lines. 
In addition to the self-consistent mean-field
potential, pairing correlations have to be included in order to
describe ground-state properties of open-shell nuclei.
Exotic nuclei, in particular, necessitate a unified description of
mean-field and pairing correlations:
the relativistic Hartree-Bogoliubov model~\cite{LVR.99}.

In addition to ground state properties, collective excitations have
been described in the time-dependent relativistic mean-field
model~\cite{Vre.95,Vre.97,Vre.99} and the 
relativistic random phase approximation (RRPA)~\cite{lhu.89,Daw.90,Rrpa.00,Ma.00}.
The RRPA represents the small amplitude limit of the
time-dependent relativistic mean-field theory. Self-consistency
will therefore ensure that the same correlations which
define the ground-state properties, also determine
the behavior of small deviations from the equilibrium.
The same effective Lagrangian generates the Dirac-Hartree
single-particle spectrum and the residual particle-hole
interaction. In most applications we have
used  the NL3 interaction \cite{LKR.97} for the effective
Lagrangian. Properties calculated with the NL3 indicate that this is probably
the best effective interaction so far, both for nuclei at and away from the
line of $\beta $-stability. This effective interaction will be also
used in the present analysis of isovector dipole response in neutron
rich nuclei. The RRPA equations form read~\cite{Daw.90}
\begin{equation}
\label{rrpaeq}
\left(
\begin{array}{cc}
A^J & B^J \\
B^{^\ast J} & A^{^\ast J}
\end{array}
\right)
\left(
\begin{array}{c}
X^{\nu,JM}_{\tilde p h} \\
Y^{\nu,JM}_{\tilde p h}
\end{array}
\right) =\omega_{\nu}\left(
\begin{array}{cc}
1 & 0 \\
0 & -1
\end{array}
\right)
\left( 
\begin{array}{c}
X^{\nu,JM}_{\tilde p h} \\
Y^{\nu,JM}_{\tilde p h}
\end{array}
\right)
\label{RRPA}
\end{equation}
The matrix elements contain the single-particle energies
and the two-body interaction
\begin{equation}
A^J_{j_{\tilde p} j_h, j_{\tilde q} j_i}
= \left(\epsilon_{j_{\tilde p}} - \epsilon_{j_{h}} \right)
\delta_{j_{\tilde p} j_{\tilde q}}\delta_{j_h j_i}
+V^J_{j_{\tilde p} j_i j_h j_{\tilde q}}
\end{equation}
\begin{equation}
B^J_{j_{\tilde p} j_h, j_{\tilde q} j_i}
= (-1)^{j_{\tilde q} - j_i + J}V^J_{j_{\tilde p} j_{\tilde q} j_h j_i}.
\end{equation}
$\tilde p$ denotes both particle and antiparticle states, 
$h$ denotes states in the Fermi sea.
An RRPA calculation, consistent with the mean-field 
model in the $no-sea$ approximation, necessitates configuration
spaces that include both particle-hole pairs and pairs formed
from occupied states and negative-energy states~\cite{Daw.90}.
The contributions
from configurations built from occupied positive-energy states and
negative-energy states are essential for current conservation and
the decoupling of the spurious state. In addition, configurations
which include negative-energy states give an important contribution
to the collectivity of excited states. In a recent study~\cite{Ma.00}
we have shown that, in order to reproduce results of time-dependent
relativistic mean-field calculations for giant resonances, the
RRPA configuration space must contain negative-energy Dirac states,
and the two-body matrix elements must include contributions
from the spatial components of the vector meson fields.

The solutions of the RRPA equations (\ref{rrpaeq}) are used to
evaluate the electric dipole response
\begin{equation} 
B(EJ,\omega_{\nu}) = \frac{1}{2J+1} 
\left| \sum_{j_{\tilde p} j_h} X^{\nu, J0}_{j_{\tilde p} j_h} \langle
j_{\tilde p} || \hat{Q}_J || j_h \rangle \\
+~(-)^{j_{\tilde p}-j_h+J} \, Y^{\nu, J0}_{j_{\tilde p} j_h}
\, \langle j_h || \hat{Q}_J || j_{\tilde p} \rangle \, \right|^2 \quad.
\label{strength} 
\end{equation}
for the isovector dipole operator
\begin{equation}
\hat{Q}_{1 \mu}^{T=1} \ = \frac{N}{N+Z}\sum^{Z}_{p=1} r_{p}Y_{1 \mu}
- \frac{Z}{N+Z}\sum^{N}_{n=1} r_{n}Y_{1 \mu}.
\label{dipop}
\end{equation}
A large configuration space is used in order to eliminate the 
spurious component from the physical states~\cite{Daw.90}.

The effect of the Dirac sea states on the isovector dipole strength 
distribution is not so pronounced as in the isoscalar case~\cite{Ma.00}.
Without the inclusion of the Dirac sea states the position of 
an isoscalar giant resonances is lowered in energy by 
several MeV. In contrast, in Fig.~\ref{figA} we illustrate
the effect of Dirac sea states on the 
isovector dipole strength distribution of $^{208}$Pb.
The solid and long-dashed curves are the RRPA
strengths with and without the inclusion of Dirac sea states, respectively.
The dotted, dot-dashed and short-dashed 
curves correspond to calculations in which
only the $\sigma$, the $\omega$ and the                               
$\rho$ meson field are included in the coupling
between the Fermi sea and Dirac sea states, respectively. 
We notice that, although the position of 
the peak is not sensitive to the inclusion of negative energy states,
these configurations affect the total intensity, i.e. the 
values of the calculated energy weighted moments. In particular, if the
negative energy Dirac sea states are not included in the RRPA 
configuration space, only 72.8\% of the energy weighted sum rule
is exhausted, as compared with the full RRPA calculation 
with both positive and negative energy states.
When only the $\sigma$, the $\omega$, or the $\rho$ meson couple 
the particle-hole states with negative energy Dirac sea
states, 89.4\%, 98.3\% and 79.2\% 
of the full RRPA energy weighted sum rule is exhausted, respectively.
The self-consistent RRPA framework used in the present
analysis includes configurations built from occupied positive-energy
states and negative-energy states.

\section{Low-lying dipole strength in oxygen and calcium isotopes}

\indent In this section the RRPA is applied in the analysis of the low-lying
isovector dipole strength in neutron rich O and Ca isotopes.
The discrete spectra are averaged with the 
Lorentzian distribution
\begin{equation} 
\label{diplor}
R\left( E \right) = \sum_{i}B(E1,1_i \rightarrow 0_f)
\frac{\Gamma^2}{4\left(E-E_{i}\right)^2-\Gamma^2},
\end{equation}
with the B(E1) values (~\ref{strength}) calculated for the 
isovector dipole operator (~\ref{dipop}), and $\Gamma = 0.5$ MeV
is an arbitrary choice for the width of the Lorentzian.
The energy of the resonance is defined as the centroid energy
\begin{equation}
\label{meanen}
\bar E = \frac{m_1}{m_0}~, 
\end{equation}
with the energy weighted moments for discrete spectra 
\begin{equation}
m_k=\sum_{i}B(E1,1_i \rightarrow 0_f) E^k_{i}.
\label{m1}
\end{equation} 
For $k=1$ this equation defines the energy weighted
sum rule (EWSR). In the present analysis the EWSR is evaluated in the
interval below 50 MeV excitation energy.

In Fig.~\ref{figB} we display the isovector dipole strength distributions
(\ref{diplor}) for $^{16}$O, $^{22}$O, $^{24}$O and the 
hypothetical nucleus $^{28}$O. Although $^{24}$O is the last 
bound oxygen isotope, in many mean-field calculations, including the 
present with the NL3 effective interaction, the neutron drip line
is located at $^{28}$O. Already for $^{16}$O the isovector dipole
strength distribution is strongly fragmented with the centroid 
energy at $\bar E$=21.8 MeV.
The thin dashed line tentatively separates the region of giant resonances
from the low-energy region below 10 MeV.
By increasing the number of neutrons, two main effects are observed:
a) an increased fragmentation of the dipole 
strength, and b) the appearance of low lying strength below 10 MeV.
The relative contribution of the low-energy region increases with the
neutron excess. This is shown in Fig.~\ref{figC} where we plot 
the ratios of the energy weighted m$_1$ moments calculated in the 
low (E$\leq$10 MeV) and high (E$>$10 MeV) energy regions, as
function of the neutron excess $N-N_{c}$, with $N_{c}=Z$.
In the extreme case of $^{28}$O, this ratio is almost 0.15, i.e.
a considerable portion of the strength function is located below 10 MeV.
A similar result for $^{28}$O was also obtained in the non-relativistic
Hartree-Fock plus RPA framework with Skyrme effective interactions~\cite{Cat.97}.
Several peaks were calculated in the region between 6 and 10 MeV. 
In the present analysis for
$^{28}$O the most collective RRPA peak at 15.2 MeV exhausts 24\% of
the EWSR. In the corresponding HF+RPA calculation of Ref.~\cite{Cat.97}, 
the most collective isovector state exhausts approximately 
15\% of the total EWSR.

In a recent experimental investigation, low-lying dipole strength,
exhausting around 5\% of the TRK sum rule for energies up to 5 MeV above
threshold, was observed for $^{18}$O, $^{20}$O and $^{22}$O~\cite{Aum.99}.
Further experimental study of the drip line nucleus $^{24}$O is planned in
the near future. In the present RRPA calculation we find 
2.5\%, 7.0\% and 8.6\% of the EWSR in the energy region below 10 MeV
for $^{22}$O, $^{24}$O and $^{28}$O, respectively. In comparison, 
the large scale shell model calculation of Ref.~\cite{SS.99} predicts 
that the low-lying dipole strength below 15 MeV exhausts 10\% of the
classical sum rule in $^{22}$O, and 8.6\% in $^{24}$O.

What is the nature of these isovector dipole states? The question whether 
the soft, i.e. low-lying dipole excitations are collective or single-particle 
has been addressed, for example, in Ref.~\cite{Sag.95} for the light
neutron halo nuclei $^{11}$Li and $^{11}$Be. It has been shown that the soft
modes, which result from the large spatial extension of the bound
single-particle states, represent a new type of non-resonant independent 
single-particle excitations. The narrow width and the large transition strength,
which characterize these excitations, are not caused by a coherent superposition
of particle-hole ({\it ph}) configurations like in collective states.

In the present RRPA calculation of the oxygen isotopes, we analyze in more 
detail the structure of the main peaks in the low-energy region of the 
isovector dipole strength distribution (\ref{figA}).
For a state at energy $\omega_{\nu}$, the contribution of a particular 
proton or neutron {\it ph} configuration is determined by the RRPA 
amplitude 
\begin{equation}
\xi_{\tilde{p} h}=\left|X^{\nu}_{\tilde{p} h}\right|^2-
\left|Y^{\nu}_{\tilde{p} h}\right|^2~,
\end{equation} 
with X and Y defined by the RRPA equation (\ref{RRPA}), and the normalization 
condition
\begin{equation}
\sum_{\tilde{p} h}\xi_{\tilde{p} h}=1~.
\label{norm}
\end{equation}

For $^{22}$O we find only one strong peak below 10 MeV: the state at
9.3 MeV exhausts 2.5\% of the EWSR, and its wave function is very simple:
single neutron excitations 
$(93\%\ 1d_{5/2} \to 2p_{3/2})$ and $(3\%\ 1d_{5/2}
\to 1f_{7/2})$. The neutron {\it ph} excitations determine also 
the main peaks in the low-energy region of $^{24}$O and $^{28}$O.
For $^{24}$O we find three strong peaks at 
6.9 MeV (3.1\% EWSR), 7.4 MeV (1.6\% EWSR) and 9.3 MeV (2.3\% EWSR).
These states correspond to the neutron {\it ph} excitations:
$(93\%\ 2s_{1/2} \to 2p_{3/2})$, 
$(96\%\ 2s_{1/2} \to 2p_{1/2})$ and $(94\%\ 1d_{5/2} \to 2p_{3/2})$,
respectively. The strength function below 10 MeV is more fragmented 
for $^{28}$O. The main neutron {\it ph} components of the states 
at 4.2 MeV, 4.9 MeV, 7.3 MeV and 8.9 MeV, are displayed in Table~\ref{TabA}.
We conclude that in all neutron rich oxygen isotopes the isovector dipole
response in the low-energy region below 10 MeV is characterized by 
single particle transitions, in contrast to the coherent superposition of 
many {\it ph} configurations, which characterizes the excitations in the 
region of giant resonances. 

The difference in the structure of isovector dipole states in the region 
below 10 Mev and in the region 
of giant resonances is illustrated in Fig.~\ref{figD}, where we display 
the transition densities for the peaks at 9.3 MeV and 20.9 MeV in $^{22}$O,
and for the peaks at 7.3 MeV and 18.1 MeV in $^{28}$O. The proton and 
neutron contributions are plotted separately; the dotted lines denote the 
isovector transition densities, and solid lines are used for the isoscalar
transition densities. As it has been also shown in Ref.~\cite{Cat.97}, 
although the isoscalar B(E1) to all states must vanish identically,
the corresponding isoscalar transition densities to individual states
need not be identically zero. The transition densities for the states 
at 20.9 MeV in $^{22}$O and at 18.1 MeV in $^{28}$O display the radial 
dependence characteristic for the isovector giant dipole resonance: 
the proton and neutron densities oscillate with opposite phases;
the amplitude of the isovector transition density is much larger than 
that of the isoscalar component; and at large radii both the isovector
and isoscalar transition densities have a similar radial dependence.
We notice that the large neutron component in the surface region 
contributes to the formation of a node for the isoscalar transition density.
In Ref.~\cite{Cat.97} it has been shown that this effect is characteristic
for neutron rich nuclei.

The transition densities for states in the low-energy region 
(the states at 9.3 MeV in $^{22}$O and at 7.3 MeV in $^{28}$O) exhibit 
a rather different radial dependence: the proton and neutron densities
in the interior region are not out of phase; there is almost no 
contribution from the protons in the surface region; the 
isoscalar transition density dominates over the isovector one in the 
interior; the neutron transition density displays a long tail in the 
radial coordinate as compared to the transition densities of the 
giant resonance states. We notice that a similar behavior of transition 
densities has been predicted for $^{6}$He, $^{11}$Li and $^{12}$Be
in Ref.~\cite{Sag.96}. It has been shown that the long tails of the 
wave functions of the loosely-bound neutrons are responsible for the 
different radial dependence of the soft low-energy states and the 
giant resonances. 

The isovector dipole strength distributions 
for $^{40}$Ca, $^{48}$Ca, $^{54}$Ca and $^{60}$Ca nuclei are plotted
in Fig.~\ref{figE}. As the neutron excess increases, the spectra 
become more fragmented, and starting with $^{54}$Ca, the onset 
of low-lying dipole strength is observed below 10 MeV. This result is
in agreement with the non-relativistic HF+RPA calculations of Ref.~\cite{Cat.97}.
Notice that no dipole strength is found below 10 MeV for $^{48}$Ca.
In the RRPA strength function of the extremely neutron rich nucleus
$^{60}$Ca, the low-energy region is strongly fragmented, with many 
peaks of similar intensity. Together, they exhaust 
10\% of the EWSR, compared to the 40\% of the EWSR exhausted by the 
main IV GDR peak at 15.2 MeV. The neutron {\it ph} configurations with 
the largest amplitudes in the RRPA wave functions of several low-lying dipole
states in $^{60}$Ca are listed in Table~\ref{TabCa}. We notice that one, 
or at most two neutron {\it ph} configurations determine the structure of the 
low-energy peaks. There is practically no contribution from proton {\it ph}
excitations. This structure is very different from that of the GDR peak, 
which is characterized by a coherent superposition of many {\it ph}
configurations. The largest single neutron {\it ph} configuration
contributes less than 20\% of the total intensity, and the ratio of the 
neutron to proton contribution 61.8\%/36.7\%=1.7 is close to the
value N/Z, as expected for a IV GDR state.

The RRPA transition densities for the state at 7.3 MeV (illustrative 
for the low-energy region) and for the GDR state at at 15.2 MeV, are 
compared in Fig.~\ref{figF}. The transition densities display the 
radial dependence and the differences that we have already discussed
above for the oxygen isotopes. In particular, we notice the long tail
of the neutron transition density for the state at 7.3 MeV. 

Our result for $^{48}$Ca, i.e. no dipole strength below 10 MeV, is at 
variance with the onset of soft dipole resonances in Ca isotopes, 
calculated in the framework of density functional theory~\cite{Cha.94}.
For $^{42,44,46,48}$Ca, in addition to several narrow peaks which are 
derived from single particle dipole transitions, a broad resonance was
found in the energy range between 5 MeV and 10 MeV. It was interpreted 
as evidence of a collective excitation: surface neutron density oscillating
out of phase with a stable $^{40}$Ca core. There is also recent experimental
evidence on low-energy dipole strength in $^{48}$Ca. While in experiments
with heavy ion reactions~\cite{Ott.99} no evidence for low-lying strength
was found in the comparison of $^{40}$Ca and $^{48}$Ca spectra, recent 
results~\cite{Har.00} of high resolution photon scattering experiments
indicate the onset of low-lying dipole strength in $^{48}$Ca. It was 
found that the sum of the B(E1) strength in the energy region between 5 MeV
and 10 MeV is about 10 times larger than in $^{40}$Ca.

\section{Onset of collective low-energy dipole resonances}
 
Our RRPA analysis of the evolution of the isovector dipole response
proceeds in this section towards medium heavy nuclei. The strength
distributions (\ref{diplor}) for $^{48}$Ni, $^{56}$Ni, $^{68}$Ni
and $^{78}$Ni are shown in Fig.~\ref{figG}. Already for $^{48}$Ni a 
peak is found in the low-energy region below 10 MeV. This state is
characterized by a single proton {\it ph} excitation 
$(96\%\ 1f_{7/2} \to 2d_{5/2})$, and it is not found in the spectra of 
heavier Ni isotopes. The low-energy dipole states built on valence 
neutron {\it ph} configurations appear only in $^{62}$Ni and heavier
isotopes. The relative contribution of the low-energy region
E$\leq$~10 MeV to the dipole strength distribution increases with the 
neutron excess. The ratio of energy weighted moments $m_{1,low}/m_{1,high}$
increases from 0.01 in $^{62}$Ni to 0.06 in $^{78}$Ni (see Fig.~\ref{figC}).
Similar to light nuclei, the low-energy spectra are dominated by 
single particle transitions. There is an important difference,
however. For the heavier Ni isotopes we find one dipole state in the 
low-energy region, which displays a more complex structure of the RRPA amplitude,
i.e. a coherent superposition of more than just a few neutron {\it ph} 
configurations. In the case of $^{68}$Ni, this is the state at 9 MeV 
(4.3\% EWSR). The distribution of neutron {\it ph} configurations for this 
state is included in Table~\ref{TabNiSn}. A dipole state with a similar 
structure in $^{78}$Ni is found at 8.9 MeV (4.0\% EWSR). The transition 
densities for this state are compared with those of the GDR state 
at 16.4 MeV in Fig.~\ref{figH}. We notice that the state at 8.9 MeV is 
characterized by a strong isoscalar transition density and a long tail of 
the neutron transition density which extends almost to 10 fm. 

Suzuki, Ikeda and Sato (SIS) analyzed 
the onset of pygmy dipole resonances
in neutron rich nuclei in the framework of the two-fluid (the core
fluid and the neutron excess fluid) Steinwedel-Jensen hydrodynamical
model~\cite{Suz.90}. SIS derived the following relation between the 
energy of the pygmy dipole resonance and the excitation energy of the GDR 
\begin{equation} 
\label{epig}
E_{PR} = \sqrt{\frac{Z(N-N_c)}{N(Z+N_c)}}E_{GDR},
\end{equation} 
where $N_c$ denotes the number of neutrons that form the core together with
Z protons, and N is the total number of neutrons.

In Fig.~\ref{figI} (upper panel) we compare the RRPA results for 
the centroid energies of the GDR states in the Ni isotopes with the 
empirical relation $E=78A^{-1/3}$~\cite{Ber.81}. The centroid energies
in the low-energy region below 10 MeV are compared with the hydrodynamical
prediction for the excitation energies of pygmy resonances (SIS) in the 
lower panel (eq. (\ref{epig}) with $N_c=Z$). We notice that the RRPA
centroid energies are in fair agreement with the empirical mass dependence
of the GDR. The SIS model predicts the position of the
pygmy dipole resonance to increase with neutron excess, in contrast 
to the RRPA results which display the opposite behavior. This discrepancy
was also noted and discussed in Ref.~\cite{Cha.94}, where the 
predictions of the SIS model were compared with results obtained
in the framework of microscopic density functional theory.
  
For the present study
a very interesting sequence of medium heavy neutron rich nuclei is 
the chain of Sn isotopes with $50\leq N\leq 82$. In Fig.~\ref{figJ}
we display the isovector dipole strength distributions (\ref{diplor})
for $^{100}$Sn, $^{114}$Sn, $^{120}$Sn and $^{132}$Sn. With the 
increase of the number of neutrons, the onset of low-lying strength
below 10 MeV is observed. The dipole states in this energy region 
exhibit a structure similar to that observed in the neutron rich Ni 
isotopes: among several peaks characterized by single particle transitions,
between 7 MeV and 9 MeV a state is found with a more distributed structure of
the RRPA amplitude, exhausting approximately 2\% of the EWSR. In $^{132}$Sn,
for example, this state is calculated at 8.6 MeV and it exhausts 1.4\% of the
EWSR. The distribution of neutron {\it ph} configurations for this soft 
mode is displayed in Table~\ref{TabNiSn}. Nine neutron {\it ph} 
configurations contribute with more than 0.1\% to the total RRPA intensity.
The total contribution of proton {\it ph} excitations is only 10.4\%,
well below the ratio Z/N expected for a GDR state. We notice that also 
in the Hartree-Fock + RPA analysis of the E1 resonances in 
$^{208}$Pb~\cite{ACS.96}, it was found that for the pygmy states the 
neutron response is a factor 10 larger than the proton response,
whereas at energies corresponding to the GDR this ratio is around 1.6, 
or roughly N/Z. A similar result was also obtained in our recent RRPA
calculation of the isovector dipole response in $^{208}$Pb~\cite{Pyg.00}.
In addition to the GDR at 12.95 MeV and several low-lying single {\it ph}
states, a collective pygmy dipole state was identified at 7.29 MeV, with 
proton {\it ph} excitations contributing only 14\% to the total RRPA
intensity.

In Fig.~\ref{figK} we plot the transition densities to the two states at
8.6 MeV and 14.8 MeV in $^{132}$Sn. In the upper panel the proton, 
neutron, isoscalar and isovector components are displayed. The radial
dependence and the differences between the transition densities are 
very similar to those calculated for the pygmy dipole and 
GDR states in $^{78}$Ni (see Fig.~\ref{figH}). In the lower panel,
for the pygmy state at 8.6 MeV (c) and for the GDR state at 14.8 MeV (d),
the contributions of the excess neutrons ($50 < N \leq 82$) (solid),
and of the proton-neutron core ($Z,N \leq 50$) (dashed) are displayed separately.
By comparing with the transition densities shown in the upper panel 
of Fig.~\ref{figK}, we notice that there is practically no contribution 
from the core neutrons ($N \leq 50$). The {\it ph} excitations of core 
neutrons are, of course, at much higher energies. For the GDR state, 
therefore, the transition densities of the core nucleons and of the
excess neutrons have opposite phases (isovector mode). The absolute radial
dependence is similar, with the amplitude strongly peaked in the surface 
region. The two transition densities have the same sign for the pygmy state at 
8.6 MeV. The core contribution, however, vanishes for large $r$ and only 
oscillations of the excess neutrons are observed on the surface of $^{132}$Sn.

The GDR centroid energies of the Sn isotopes are compared with the empirical 
mass dependence $E=78A^{-1/3}$~\cite{Ber.81} in the upper panel of 
Fig.~\ref{figL}. 
In the lower panel the centroid energies of the isovector dipole strength 
in the low-energy region below 10 MeV are compared with the energies 
of the pygmy dipole states calculated in the two-fluid hydrodynamical
model (SIS)~\cite{Suz.90}. As we have already observed in the case of 
Ni isotopes, in contrast to the predicted increase of the pygmy excitation
energy with neutron excess in the SIS model, the RRPA centroid energies
decrease with the number of neutrons. 

As a final example, we analyze the isovector dipole response of a nucleus
with an extreme neutron to proton number ratio: $^{122}$Zr (Z=40, N=82).
The strength distribution is shown in the left panel of Fig.~\ref{figM}.
In addition to the GDR structure, the pygmy dipole state is identified at
7.7 MeV excitation energy. This state exhausts 3.3\% of the EWSR. 15 
neutron {\it ph} configurations contribute with more than 0.1\% to the 
total RRPA intensity of the pygmy dipole state, compared to 27 neutron 
{\it ph} configurations for the GDR state. The relative contribution of
the proton {\it ph} excitations is 10.5\% for the pygmy state, 
and 29.0\% for the GDR state. The transition densities to the 
pygmy dipole state are displayed in the right panel of Fig.~\ref{figM}.
Similar to the structure observed for the pygmy states in Ni and Sn 
isotopes, we notice a very pronounced contribution of the neutron excess
in the surface region. 

The collective dynamics of the pygmy dipole state is also exemplified in
the analysis of transition currents. In Fig.~\ref{figN} we plot 
the velocity fields for the pygmy state at 7.7 MeV excitation energy.
The velocity distributions are derived from the corresponding
transition currents, following the procedure described in Ref.~\cite{Ser.83}.
The velocity field of the proton-neutron core
($Z\leq 40, N \leq 50$) (left panel), is separated from the contribution
of the excess neutrons ($50 < N \leq 82$) (right panel). A vector of
unit length is assigned to the largest velocity. All the other
velocity vectors in both panels are normalized accordingly. The velocity fields
of the core nucleons and of the excess neutrons are in phase in the interior 
region. They have opposite phases in the surface region, though the 
largest velocities correspond to the vibrations of the excess neutrons 
on the surface. The core velocities are much smaller 
than those of the excess neutrons on the surface, in accordance with the 
transition densities shown in Fig.~\ref{figM}.

\section{Summary}

In the present work we have applied the relativistic random phase 
approximation (RRPA) in the analysis of the evolution of the
isovector dipole response in nuclei with a large neutron excess.
The relativistic mean-field model, which has been very 
successfully applied in the description of ground-state
properties of exotic nuclei far from stability, is extended
to the RRPA in order to study the possible onset of low-energy 
collective isovector modes in nuclei with extreme isospin
values. Among many new and unique structure phenomena exhibited
by nuclei with a large neutron excess, the possible occurrence
of collective isovector modes in the energy region below
the giant resonances has recently attracted considerable interest.
In particular, theoretical and experimental studies have been
recently reported on the low-energy collective isovector dipole mode,
i.e. the pygmy resonance. The isovector pygmy modes could
provide important information on the isospin and density dependence
of the effective nuclear interaction. For example, the pygmy 
dipole resonance can be directly related to the neutron excess,
and therefore the splitting between the GDR and the pygmy resonance
represents a measure of the neutron
skin. Precise information on neutron skin in heavy nuclei is essential
for the quantification of the isovector channel of effective nuclear
forces.

In the analysis of the dynamics of isovector dipole modes 
we have used the RRPA based on effective mean-field Lagrangians with 
nonlinear meson self-interaction terms. The NL3 parameter set for the
effective mean-field Lagrangian, used in the present calculations,
reproduces ground state properties not only in nuclei along the valley of
$\beta$-stability, but also in exotic nuclei with extreme isospin
values and close to the particle drip lines. The study of the
evolution of collectivity in the isovector dipole response
in the low-energy region, 
includes neutron rich isotopes of O, Ca, Ni, Zr, and Sn.
We have analyzed the isovector dipole strength distributions, the
transition densities, the neutron and proton {\it ph} excitations
which determine the structure of the RRPA transition amplitudes, 
the transition currents and velocity distributions of excess
neutron and core densities.

The dipole response in nuclei with large neutron to proton
ratio is characterized by the fragmentation of the strength distribution
and its spreading into the low-energy region, and by the mixing
of isoscalar and isovector modes. In light nuclei the onset of
dipole strength in the low-energy region is due to single particle
excitations of the loosely bound neutrons. In heavier nuclei,
low-lying dipole states appear which are characterized 
by a more distributed structure of the RRPA amplitude, 
exhausting approximately 2\% of the EWSR.
Among several peaks characterized by single particle transitions,
a single collective isovector dipole state is identified below 10 MeV. 
A coherent superposition of many neutron particle-hole configurations
characterizes its RRPA amplitude. An analysis
of the corresponding transition densities and velocity distributions
reveals the dynamics of the dipole pygmy resonance: the vibration of the
excess neutrons against the inert core composed of equal numbers of
protons and neutrons.
\newpage
{\bf Acknowledgments}

This work has been supported in part by the
Bundesministerium f\"ur Bildung und Forschung under
contract 06~TM~979, by the Deutsche Forschungsgemeinschaft, 
and by the Gesellschaft f\"ur Schwerionenforschung (GSI)
Darmstadt.  

\newpage
\begin{figure} 

\caption{Isovector dipole strength distributions in $^{208}$Pb calculated with
the NL3 effective interaction. 
The solid and long-dashed curves are the RRPA
strengths with and without the inclusion of Dirac sea states, respectively.
The dotted, dot-dashed and short-dashed curves correspond to calculations in which
only the $\sigma$, the $\omega$ and the $\rho$ meson field are included in the coupling
between the Fermi sea and Dirac sea states, respectively.}
\label{figA}
\end{figure}

\begin{figure}
\caption{RRPA isovector dipole strength distributions in oxygen isotopes.
The thin dashed line tentatively separates the region of giant resonances
from the low-energy region below 10 MeV.}
\label{figB}
\end{figure}

\begin{figure}
\caption{Ratio of the energy weighted moments m$_1$ in the low-energy 
region (E$\leq$10 MeV) and in the region of giant resonances (E$>$10 MeV).
For a number of nuclei the ratio m$_{1,LOW}$/m$_{1,HIGH}$ is plotted as
function of the number of excess neutrons.} 
\label{figC}
\end{figure}

\begin{figure}
\caption{Transition densities for the peaks at 9.3 MeV and 20.9 MeV in 
$^{22}$O, and for the peaks at 7.3 MeV and 18.1 MeV in $^{28}$O.
Both isoscalar and isovector transition densities are displayed,
as well as the separate proton and neutron contributions. All transition
densities are multiplied by $r^2$.}
\label{figD}
\end{figure}

\begin{figure}
\caption{RRPA isovector dipole strength distributions in Ca isotopes.
The thin dashed line tentatively separates the region of giant resonances
from the low-energy region below 10 MeV.}
\label{figE}
\end{figure}

\begin{figure}
\caption{Same as in Fig. \protect\ref{figD}, but for the peaks 
at 7.3 MeV and 15.2 MeV in $^{60}$Ca.}
\label{figF}
\end{figure}

\begin{figure}
\caption{Same as in Fig. \protect\ref{figE}, but for the Ni isotopes.}
\label{figG}
\end{figure}

\begin{figure}
\caption{Same as in Fig. \protect\ref{figD}, but for the peaks 
at 8.9 MeV and 16.4 MeV in $^{78}$Ni.}
\label{figH}
\end{figure}

\begin{figure}
\caption{Centroid energies(~\ref{meanen}) of the isovector dipole strength
distribution in the region of giant resonances (upper panel), and 
in the low-energy region below 10 MeV (lower panel), as functions
of the mass number of the Ni isotopes. The IV GDR centroid
energies are compared with the empirical expression $E=78A^{-1/3}$.
The energies in the lower panel are compared with the hydrodynamical
prediction~\cite{Suz.90} for the excitation energy of the pygmy resonance (SIS).}
\label{figI}
\end{figure}

\begin{figure}
\caption{RRPA isovector dipole strength distributions in Sn isotopes.
The thin dashed line tentatively separates the region of giant resonances
from the low-energy region below 10 MeV.}
\label{figJ}
\end{figure}

\begin{figure}
\caption{Isovector (IV) and isoscalar (IS) dipole transition densities
for the states at 8.6 MeV (a) and 14.8 MeV in $^{132}$Sn.
The separate proton and neutron contributions to the transition densities
are also shown. In the lower part of the figure the contributions of the excess
neutrons ($50 < N \leq 82$) (solid), and of the proton-neutron core
($Z,N \leq 50$) (dashed) are displayed separately for the state 
at 8.6 MeV (c), and 14.8 MeV (d). The transition
densities are multiplied by $r^2$.}
\label{figK}
\end{figure}

\begin{figure}
\caption{Centroid energies(~\ref{meanen}) of the isovector dipole strength
distribution in the region of giant resonances (upper panel), and 
in the low-energy region below 10 MeV (lower panel), as functions
of the mass number of the Sn isotopes. The IV GDR centroid
energies are compared with the empirical expression $E=78A^{-1/3}$.
The energies in the lower panel are compared with the hydrodynamical
prediction~\cite{Suz.90} for the excitation energy of the pygmy resonance (SIS).}
\label{figL}
\end{figure}

\begin{figure}
\caption{Isovector dipole strength distribution in $^{122}$Zr (left panel),
and transition densities for the peak at 7.7 MeV (right panel).
Both isoscalar and isovector transition densities are displayed,
as well as the separate proton and neutron contributions (panel b).
The contributions of the excess
neutrons ($50 < N \leq 82$) (solid), and of the proton-neutron core
($Z \leq 40, N \leq 50$) (dashed) are displayed separately in panel (c).}
\label{figM}
\end{figure}

\begin{figure}
\caption{Velocity distributions for the RRPA state at
7.7 MeV in $^{122}$Zr. The velocity field of the proton-neutron core
($Z \leq 40, N \leq 50$) (left panel), is separated from the contribution
of the excess neutrons ($50 < N \leq 82$) (right panel).}
\label{figN}
\end{figure}

\newpage
\begin{table}[t]
\caption{For the main peaks in the low-energy region
of the isovector dipole strength distribution in
$^{28}$O (left column), which exhaust a certain percentage
of the EWSR (middle column), the neutron particle-hole
configurations with largest amplitudes are displayed 
in the right column. The percentage assigned to a
particular $p-h$ configuration 
refers to the normalization of the RRPA amplitudes
(\protect\ref{norm}).}
\label{TabA}
\begin{center}  
\begin{tabular}{| l | l | l |}
\hline
E[MeV] & EWSR[\%] & neutron $p-h$ configuration \\
\hline
4.2 & 0.9 & $(92\%\ 1d_{3/2} \to 2p_{3/2})$  \\
\hline
4.9 & 1.4 & $(91\%\ 1d_{3/2} \to 2p_{1/2})$ \\
\ \ & \ \ & $(\ 6\%\ 1d_{3/2} \to 2p_{3/2})$ \\
\hline
7.3 & 1.9 & $(92\%\ 2s_{1/2} \to 2p_{3/2})$ \\
\hline 
8.9 & 6.3 & $(71\%\ 1d_{3/2} \to 1f_{5/2})$ \\
\ \ & \ \ & $(16\%\ 1d_{5/2} \to 1f_{7/2})$ \\
\ \ & \ \ & $(\ 3\%\ 1d_{5/2} \to 2p_{3/2})$ \\
\hline 
\end{tabular}
\end{center}
\end{table}
\begin{table}[b]
\caption{Same as in Table 1, but for the main peaks in
the low-energy region of the isovector dipole strength 
distribution in $^{60}$Ca.}
\label{TabCa}
\begin{center}  
\begin{tabular}{| l | l | l |}
\hline
E[MeV] & EWSR[\%] & neutron $p-h$ configuration \\
\hline 
5.7 & 1.2 & $(93.1\%\ 2p_{1/2} \to 3s_{1/2})$  \\
\ \ & \ \ & $(\ 3.6\%\ 1f_{5/2} \to 2d_{5/2})$ \\
\hline
6.8 & 2.6 & $(82.8\%\ 1f_{5/2} \to 2d_{3/2})$ \\
\ \ & \ \ & $(\ 6.1\%\ 1f_{5/2} \to 2d_{5/2})$ \\
\ \ & \ \ & $(\ 3.1\%\ 1f_{7/2} \to 1g_{9/2})$ \\
\ \ & \ \ & $(\ 2.3\%\ 2p_{3/2} \to 2d_{5/2})$ \\
\ \ & \ \ & $(\ 1.1\%\ 2p_{3/2} \to 3s_{1/2})$ \\
\hline
7.3 & 1.6 & $(49.3\%\ 2p_{1/2} \to 2d_{3/2})$ \\
\ \ & \ \ & $(41.3\%\ 2p_{3/2} \to 3s_{1/2})$ \\
\ \ & \ \ & $(\ 3.2\%\ 1f_{5/2} \to 2d_{3/2})$ \\
\ \ & \ \ & $(\ 1.5\%\ 2p_{3/2} \to 2d_{5/2})$ \\
\hline 
7.9 & 2.0 & $(83.8\%\ 2p_{3/2} \to 2d_{5/2})$ \\
\ \ & \ \ & $(\ 5.7\%\ 1f_{5/2} \to 2d_{3/2})$ \\
\ \ & \ \ & $(\ 3.5\%\ 1f_{7/2} \to 1g_{9/2})$ \\
\ \ & \ \ & $(\ 2.4\%\ 1f_{5/2} \to 1g_{7/2})$ \\
\hline
9.9 & 2.1 & $(57.7\%\ 1f_{5/2} \to 1g_{7/2})$ \\
\ \ & \ \ & $(22.0\%\ 1f_{7/2} \to 1g_{9/2})$ \\
\ \ & \ \ & $(\ 3.3\%\ 1f_{7/2} \to 2d_{5/2})$ \\
\hline
\end{tabular}
\end{center}
\end{table}
\newpage
\begin{table}[c]
\caption{Distribution of  neutron particle-hole configurations
for the states at 9.0 MeV (4.3\% EWSR) in $^{68}$Ni (left column),
and at 8.6 MeV state (1.4\% EWSR) in $^{132}$Sn (right column).
The percentage of a $p-h$ configuration
refers to the normalization of the RRPA amplitudes (\protect\ref{norm}). 
Only configurations which contribute more than 0.1\% are displayed.}
\label{TabNiSn}
\begin{center}  
\begin{tabular}{| l || l |}
\hline
 $^{68}$Ni at 9.0 MeV  & $^{132}$Sn at 8.6 MeV  \\
\hline 
$(26.1\%\ 1f_{5/2} \to 2d_{5/2})$ & $(28.2\%\ 2d_{3/2} \to 2f_{5/2})$ \\
$(22.9\%\ 2p_{3/2} \to 2d_{5/2})$ & $(21.9\%\ 2d_{5/2} \to 2f_{7/2})$ \\
$(11.3\%\ 1f_{7/2} \to 1g_{9/2})$ & $(19.7\%\ 2d_{3/2} \to 3p_{1/2})$ \\
$(10.3\%\ 2p_{1/2} \to 2d_{3/2})$ & $(10.5\%\ 1h_{11/2} \to 1i_{13/2})$ \\
$(10.0\%\ 1f_{5/2} \to 2d_{3/2})$ & $(\ 3.5\%\ 2d_{5/2} \to 3p_{3/2})$ \\
$(\ 8.2\%\ 2p_{3/2} \to 3s_{1/2})$ & $(\ 1.9\%\ 1g_{7/2} \to 2f_{5/2})$ \\ 
$(\ 1.4\%\ 2p_{1/2} \to 3s_{1/2})$ & $(\ 1.5\%\ 1g_{7/2} \to 1h_{9/2})$ \\
$(\ 1.0\%\ 1f_{5/2} \to 1g_{7/2})$ & $(\ 0.6\%\ 1g_{7/2} \to 2f_{7/2})$ \\
$(\ 0.3\%\ 1s_{5/2} \to 3d_{3/2})$ & $(\ 0.6\%\ 2d_{3/2} \to 3p_{3/2})$ \\
\hline
\end{tabular}
\end{center} 
\end{table}
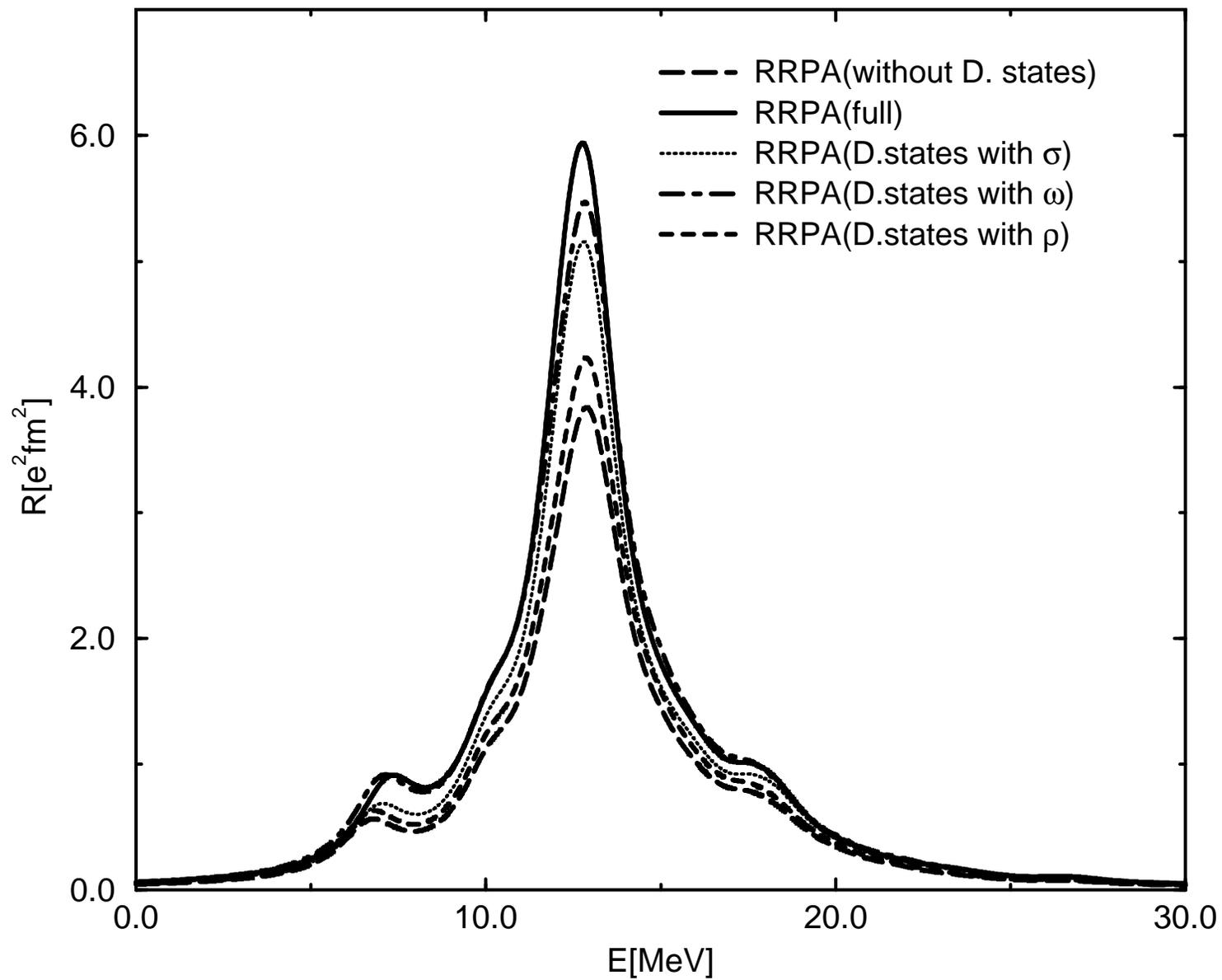
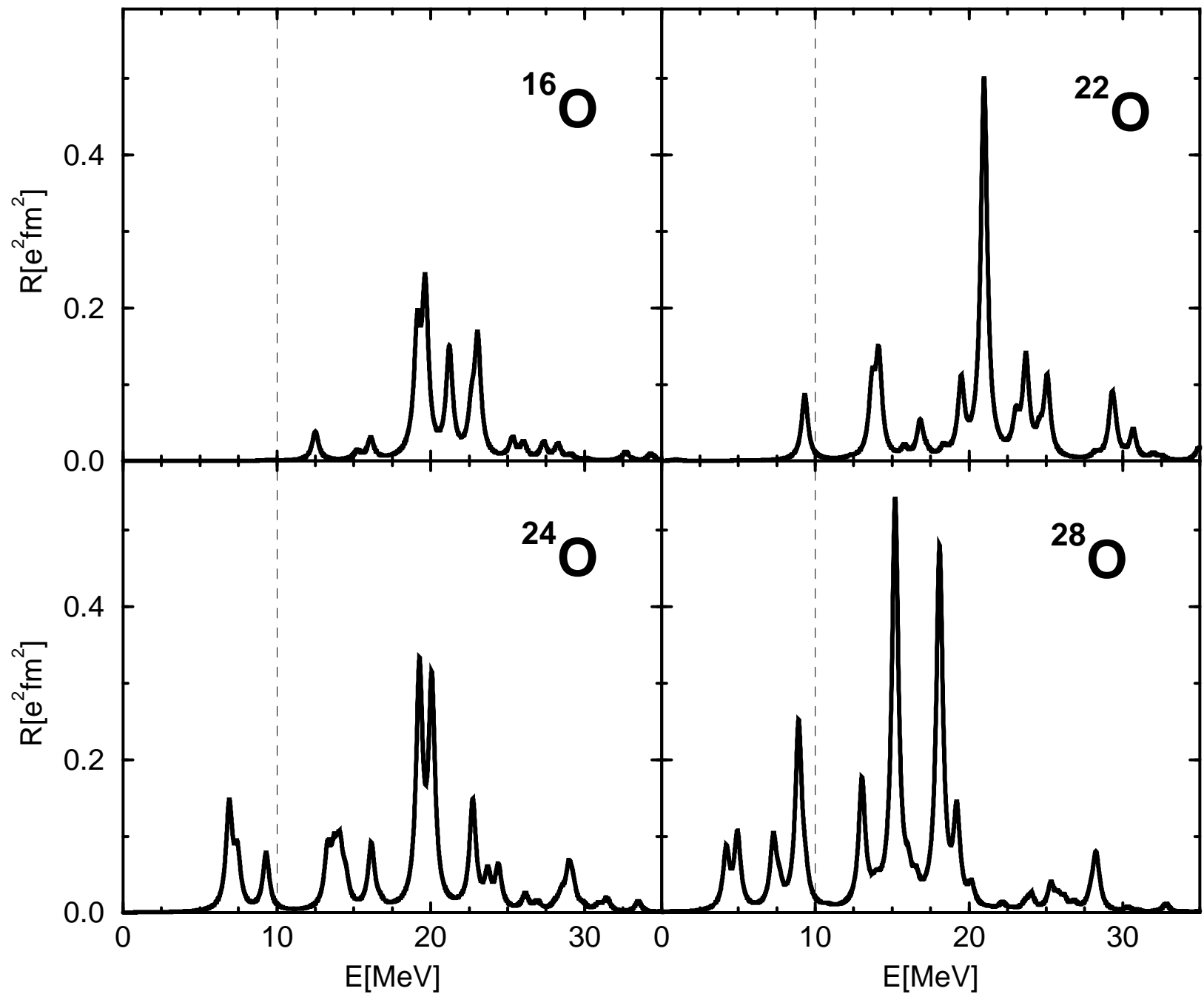
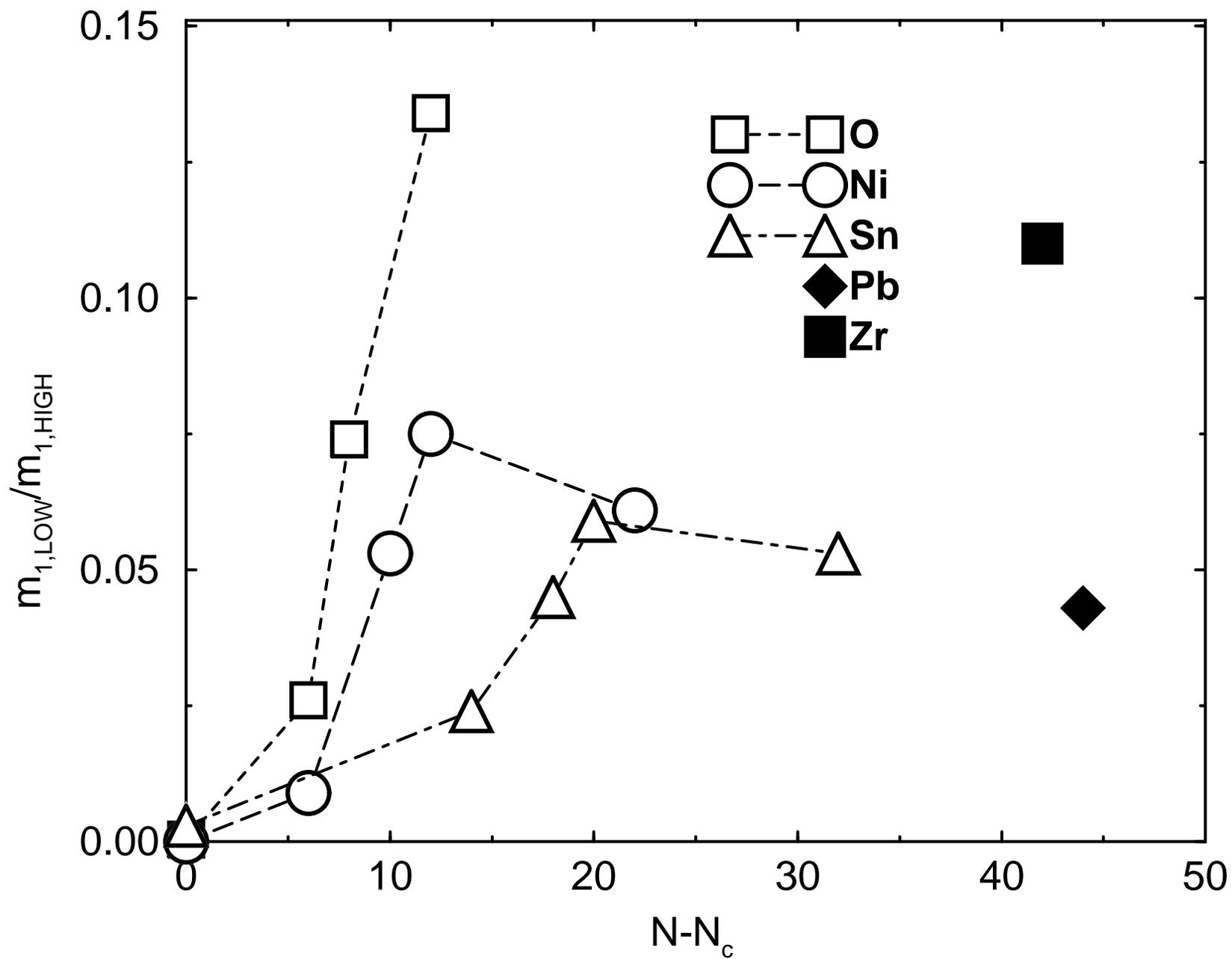
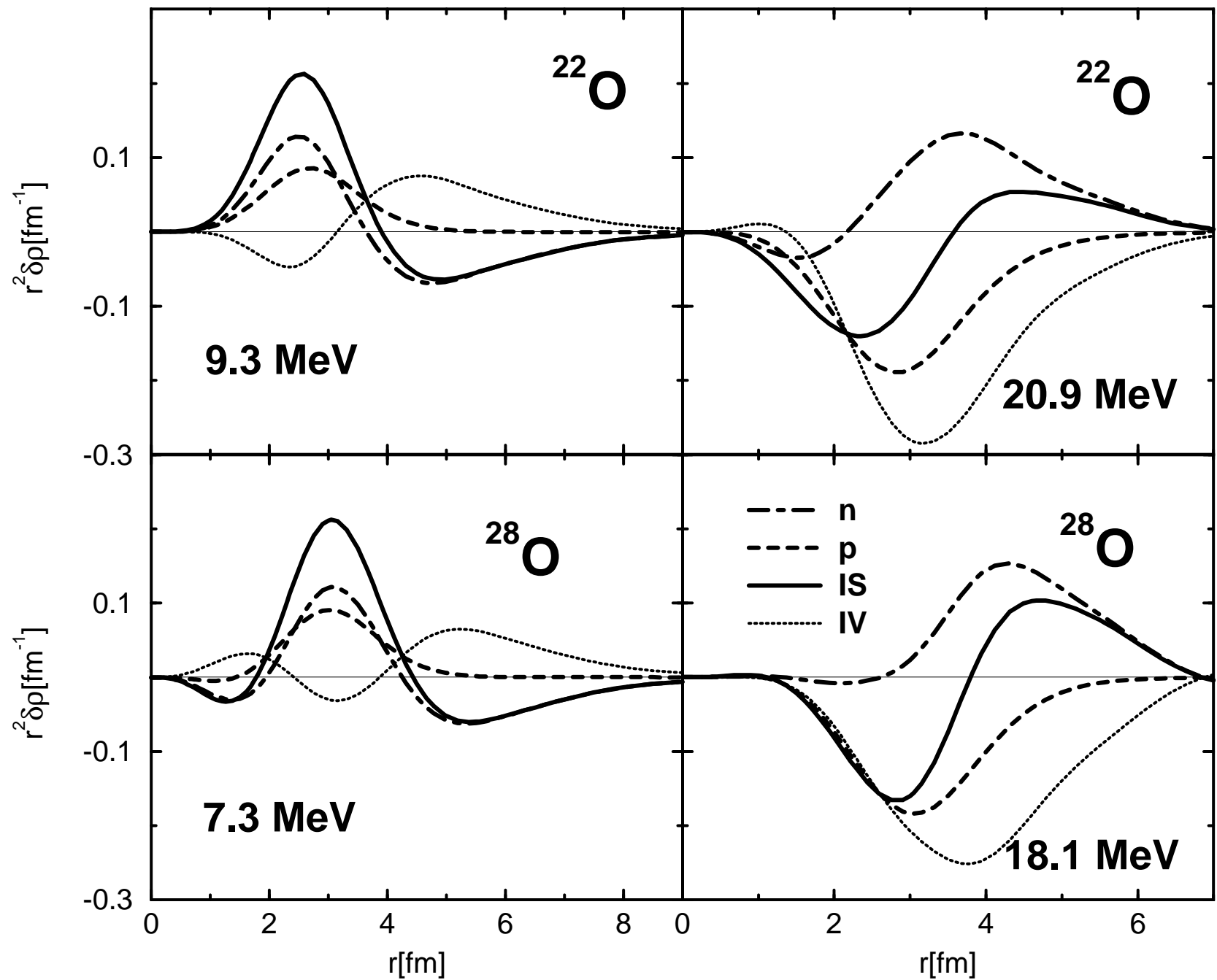
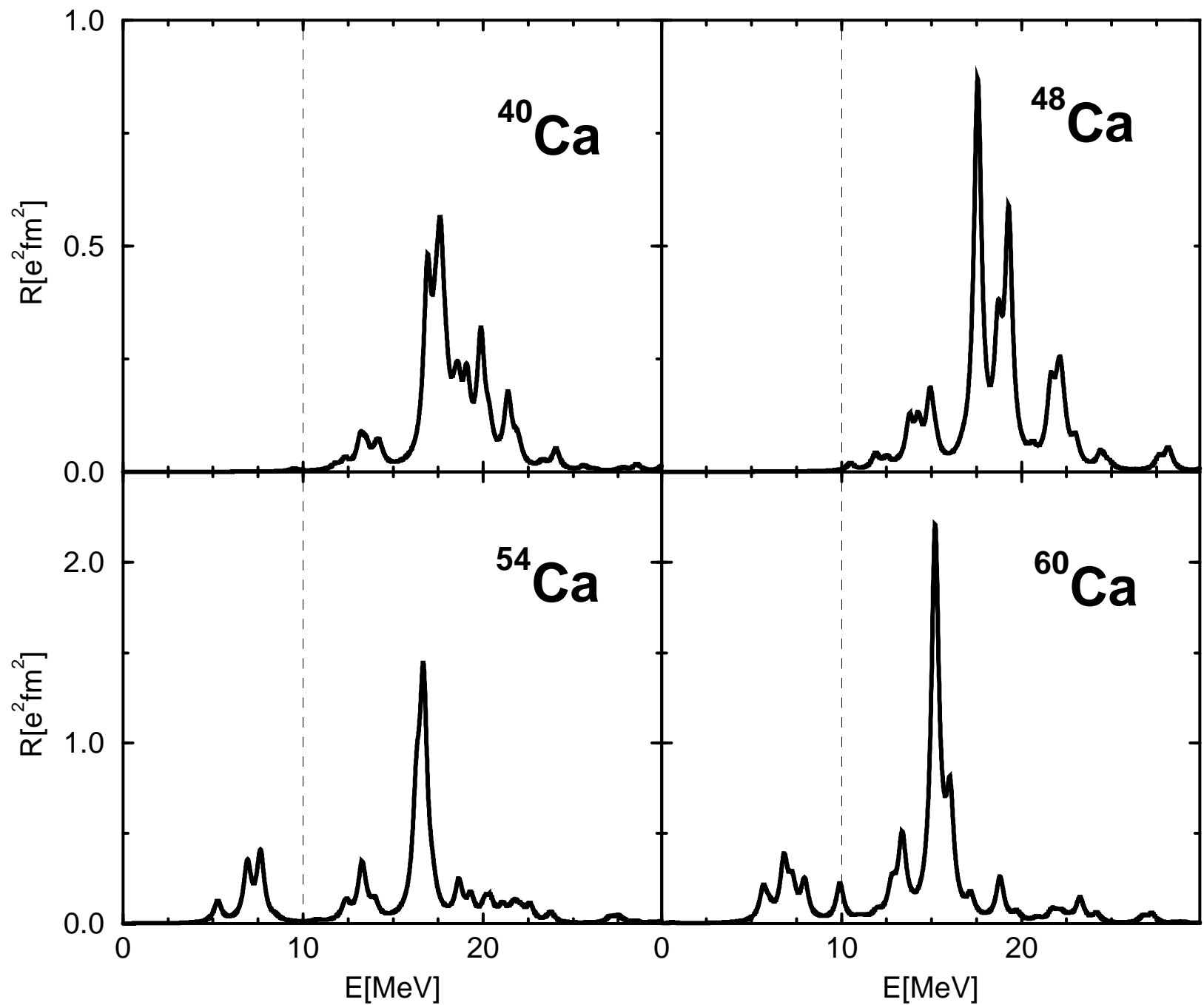
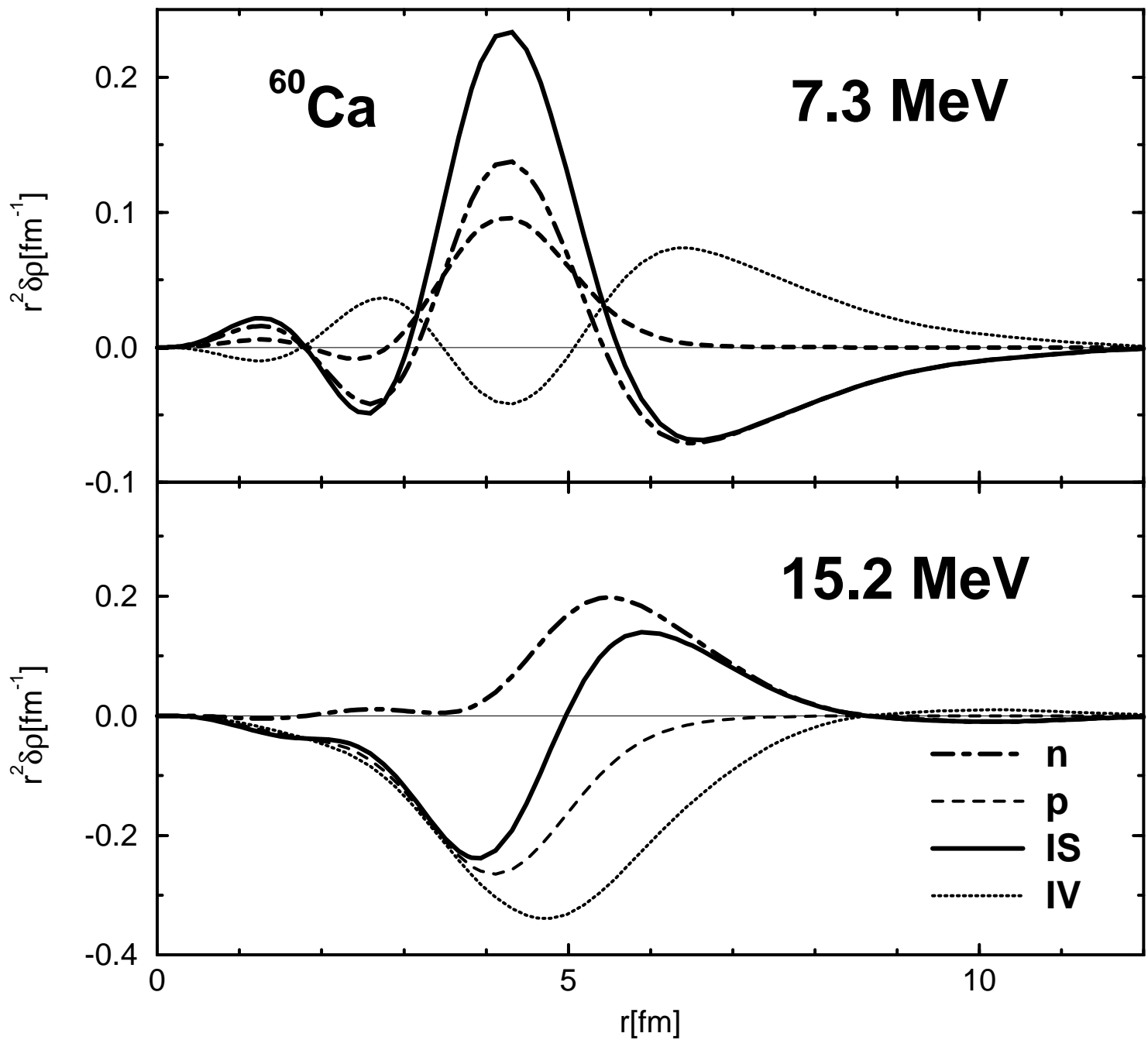
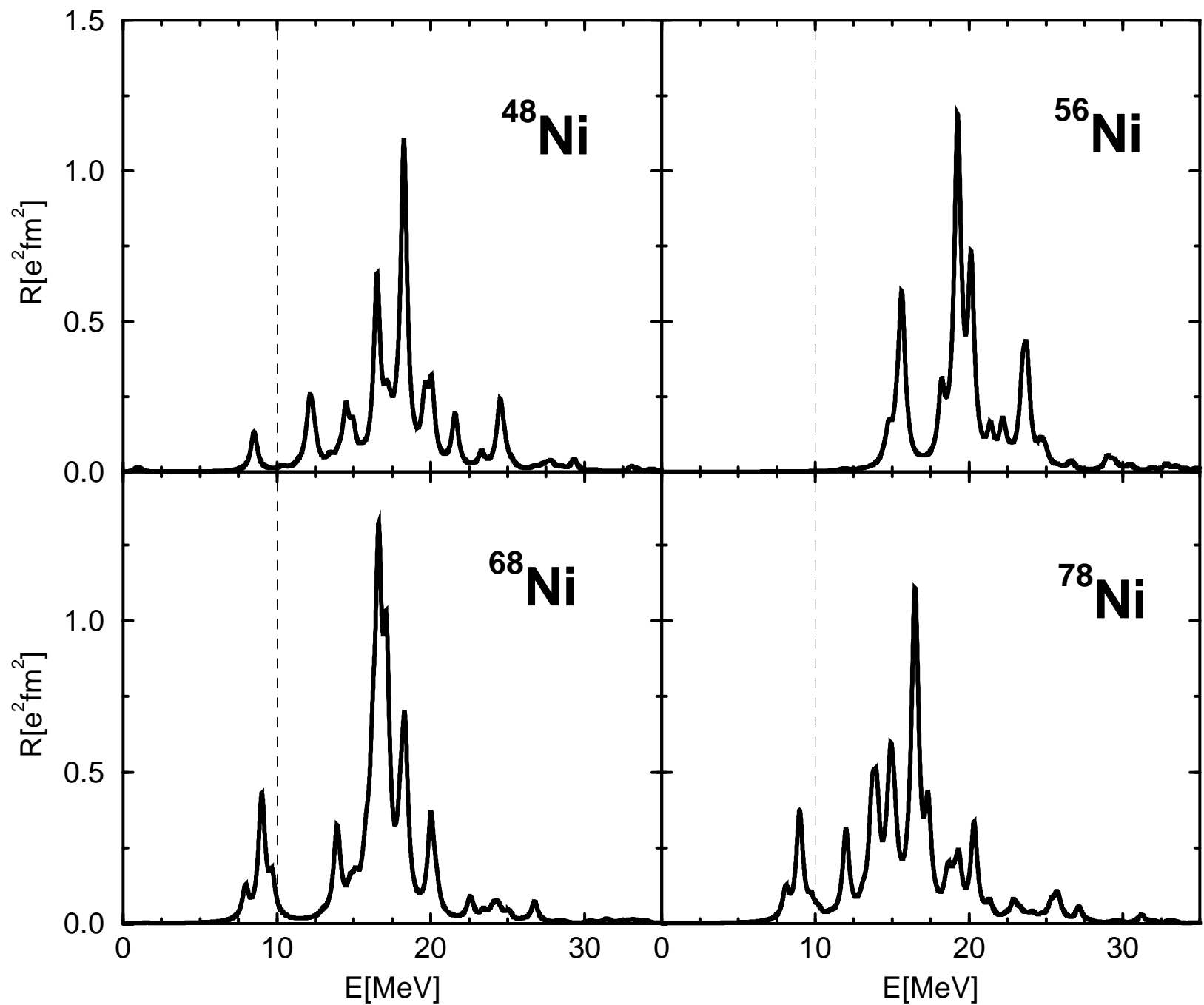
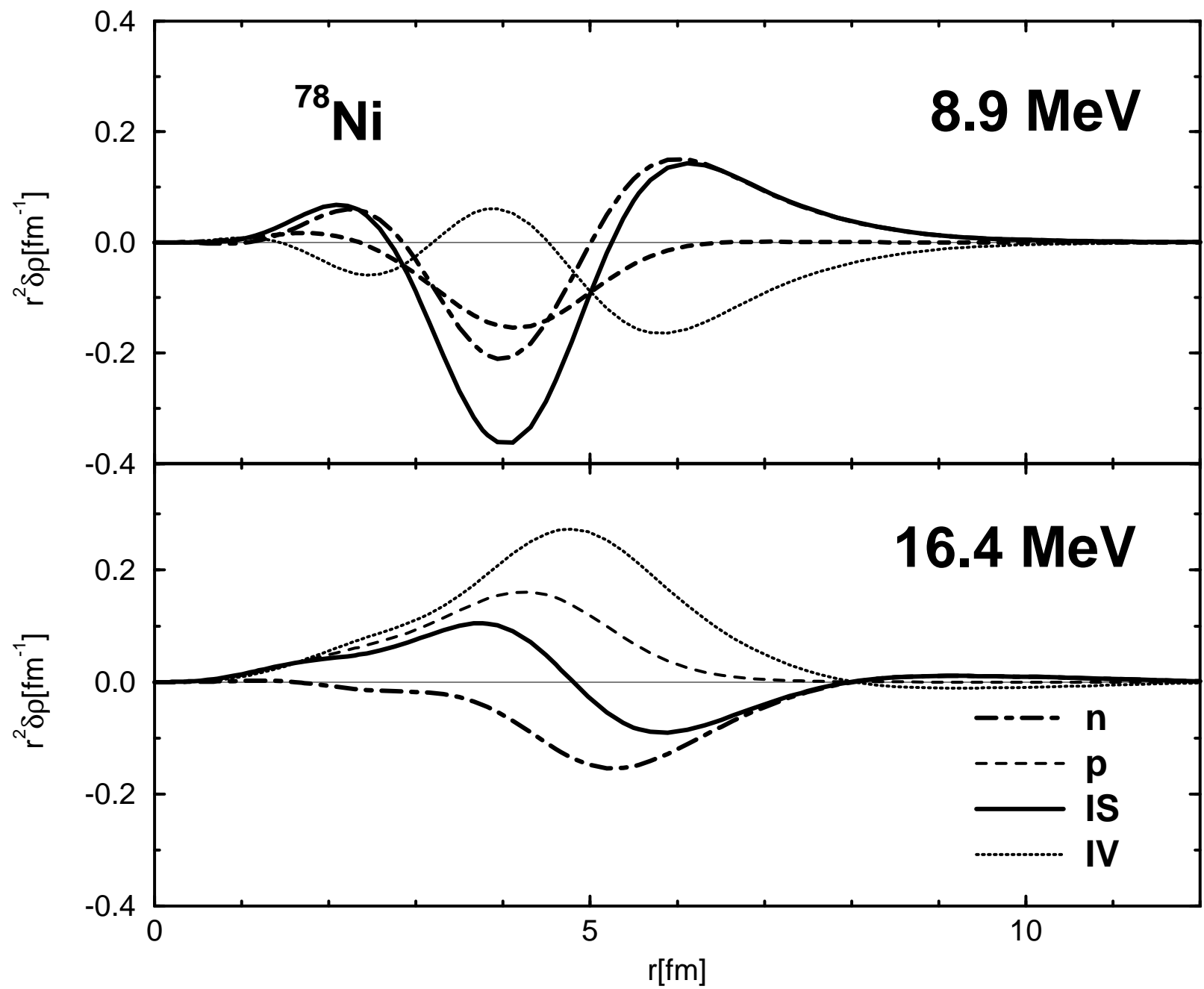
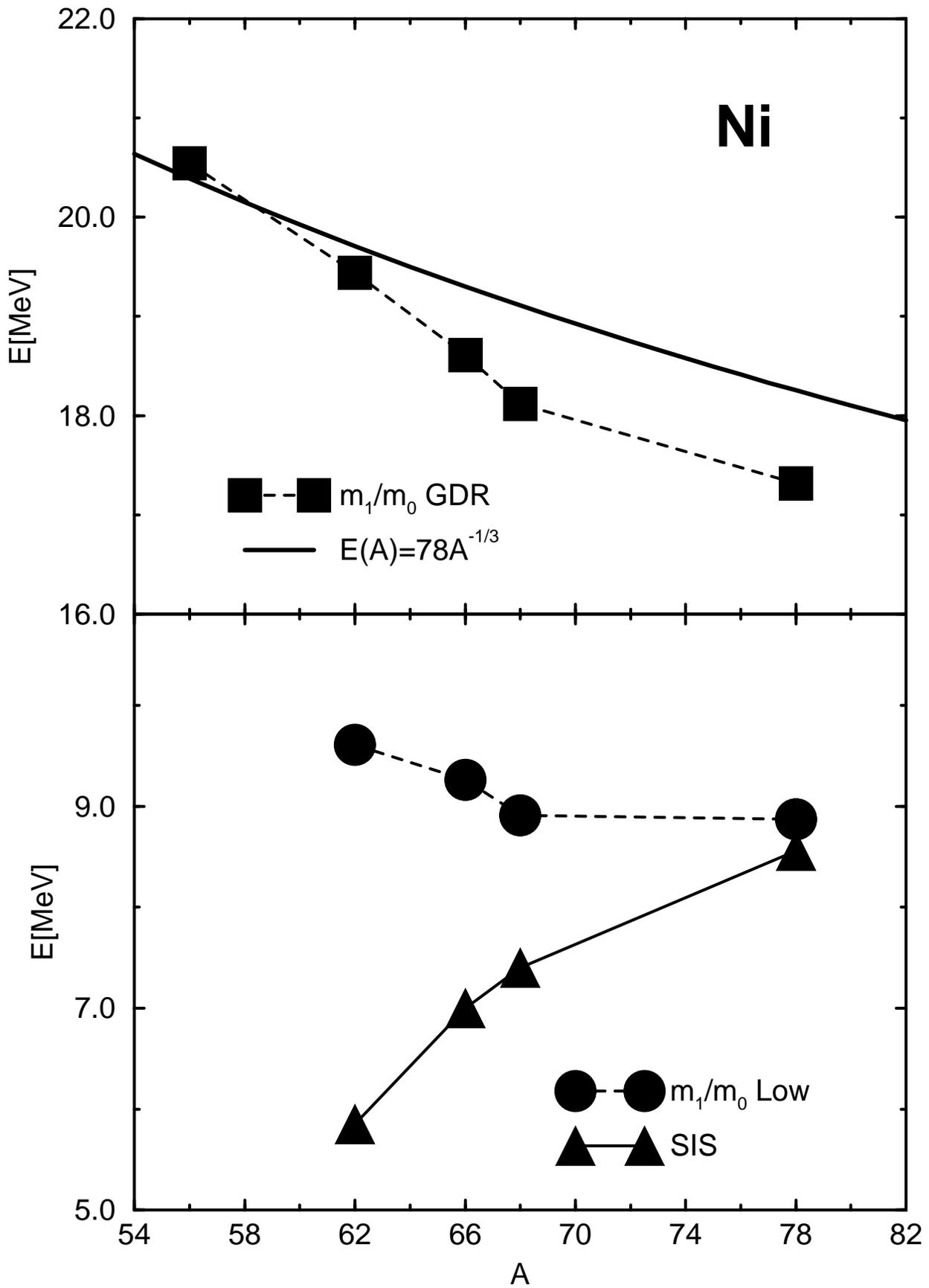
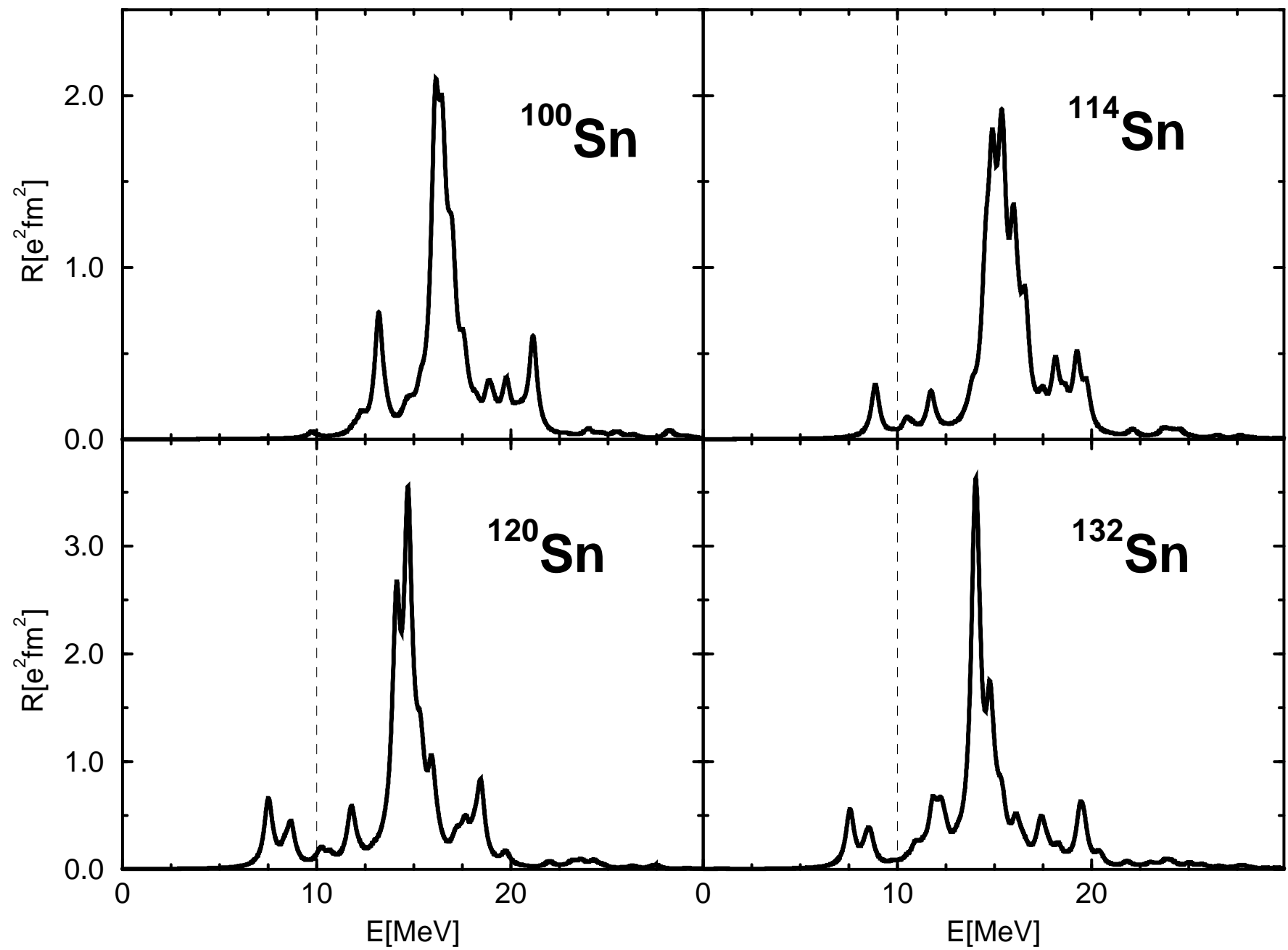
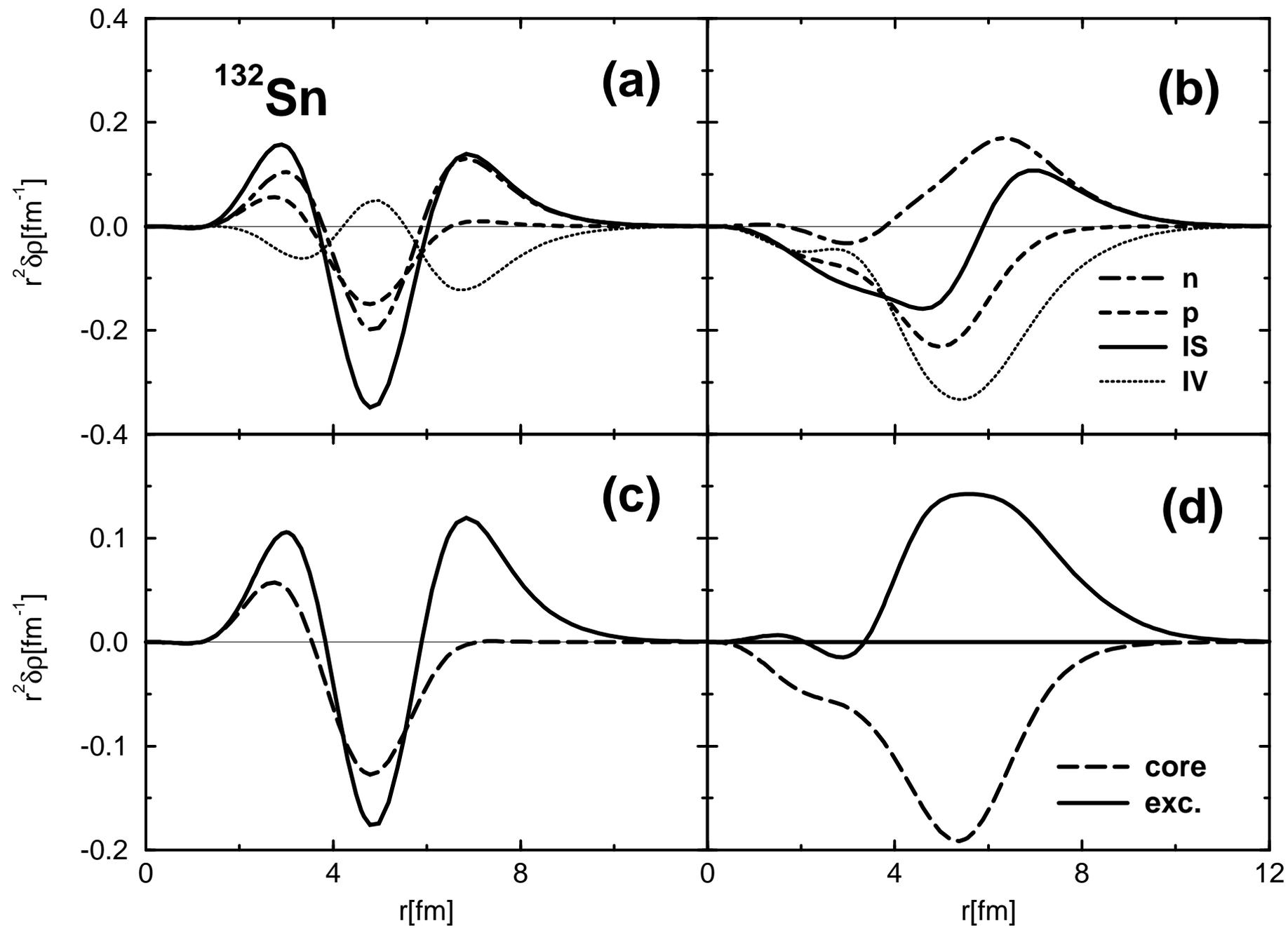
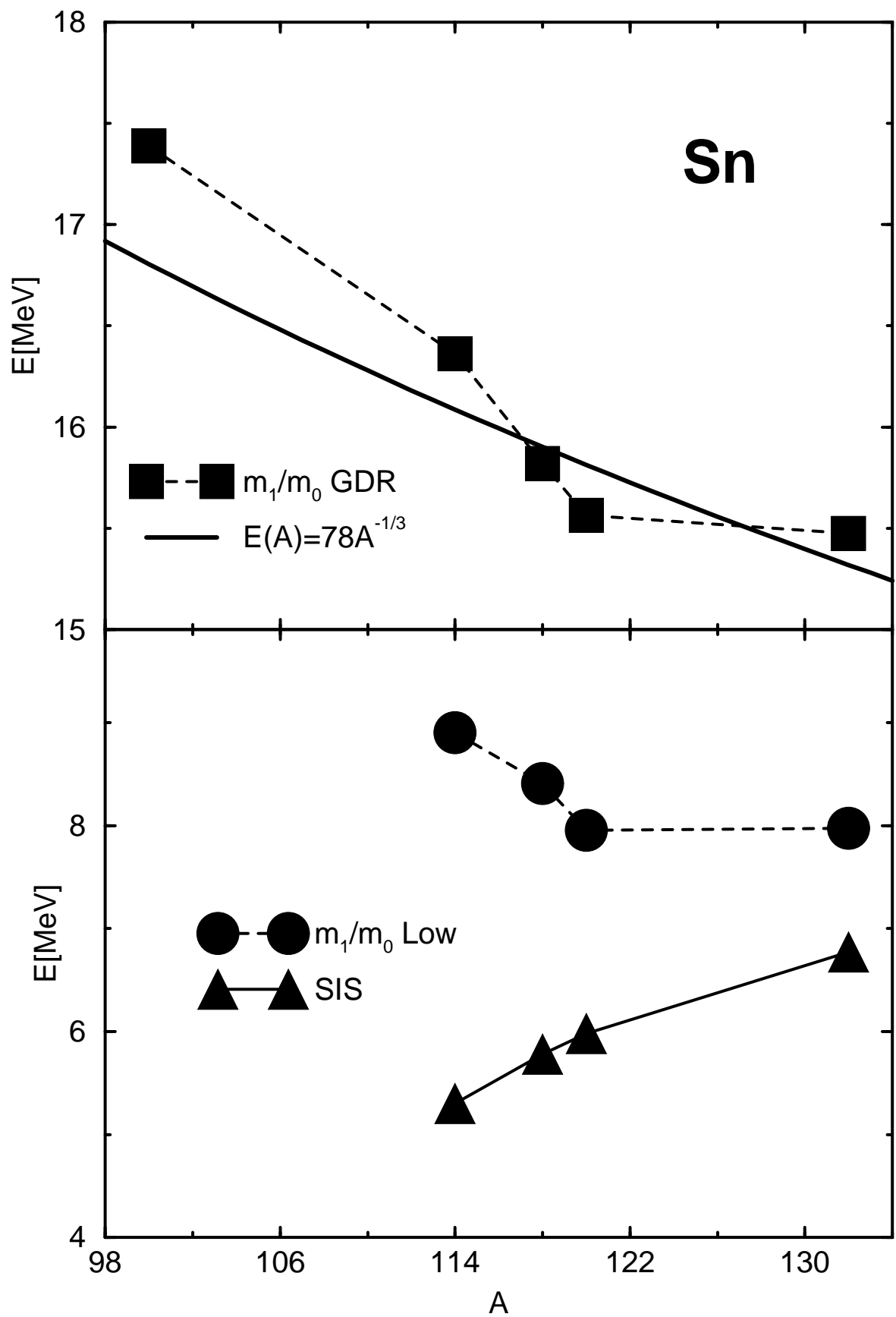
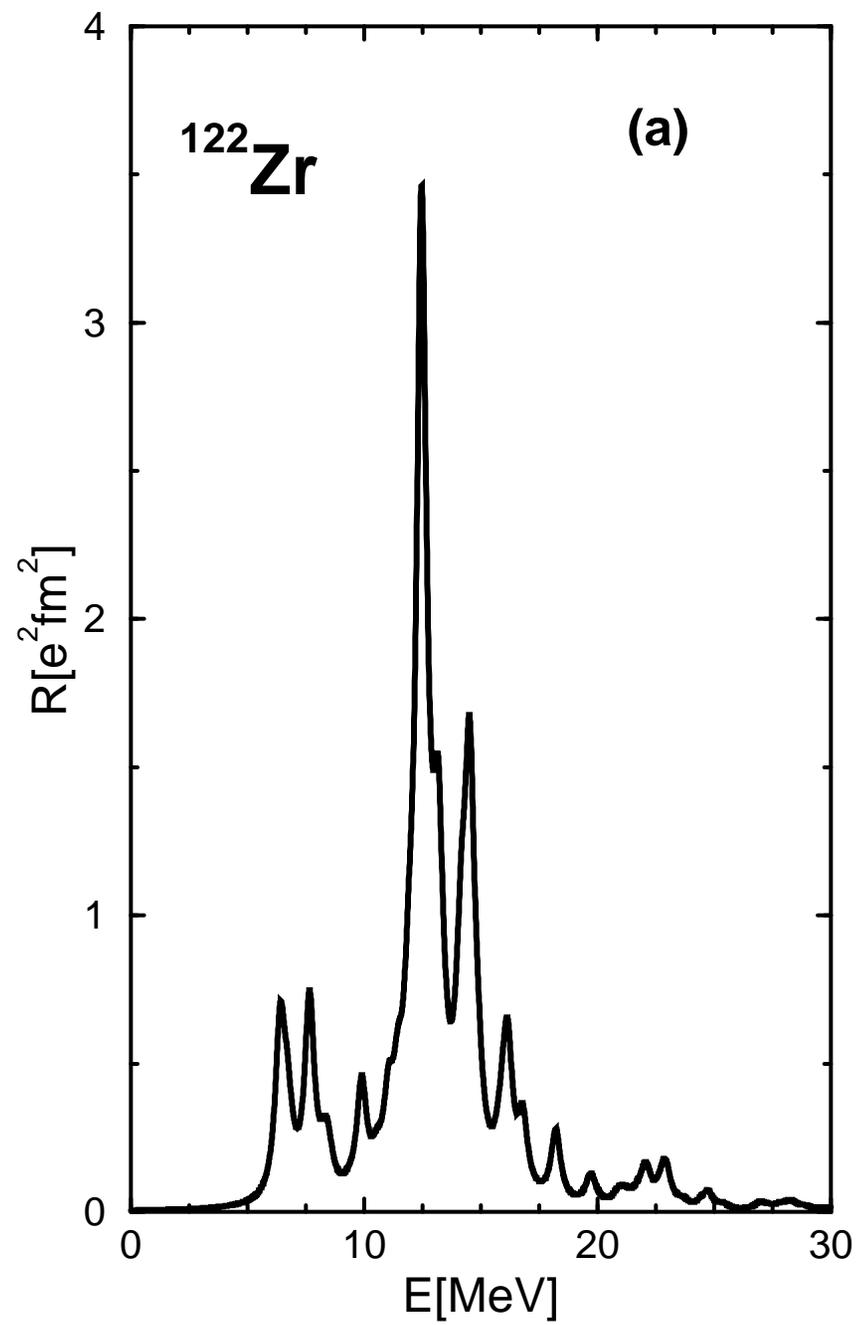
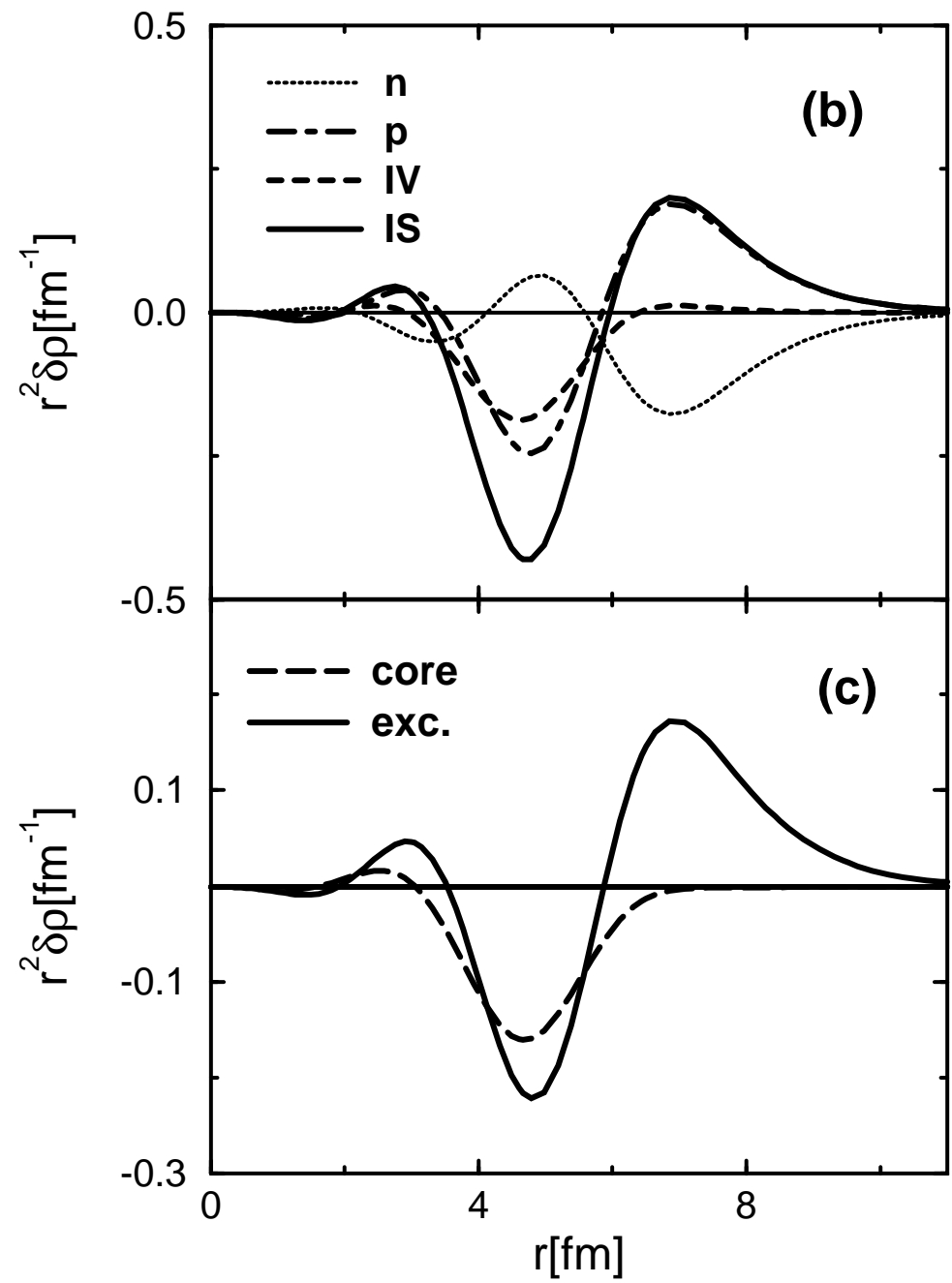
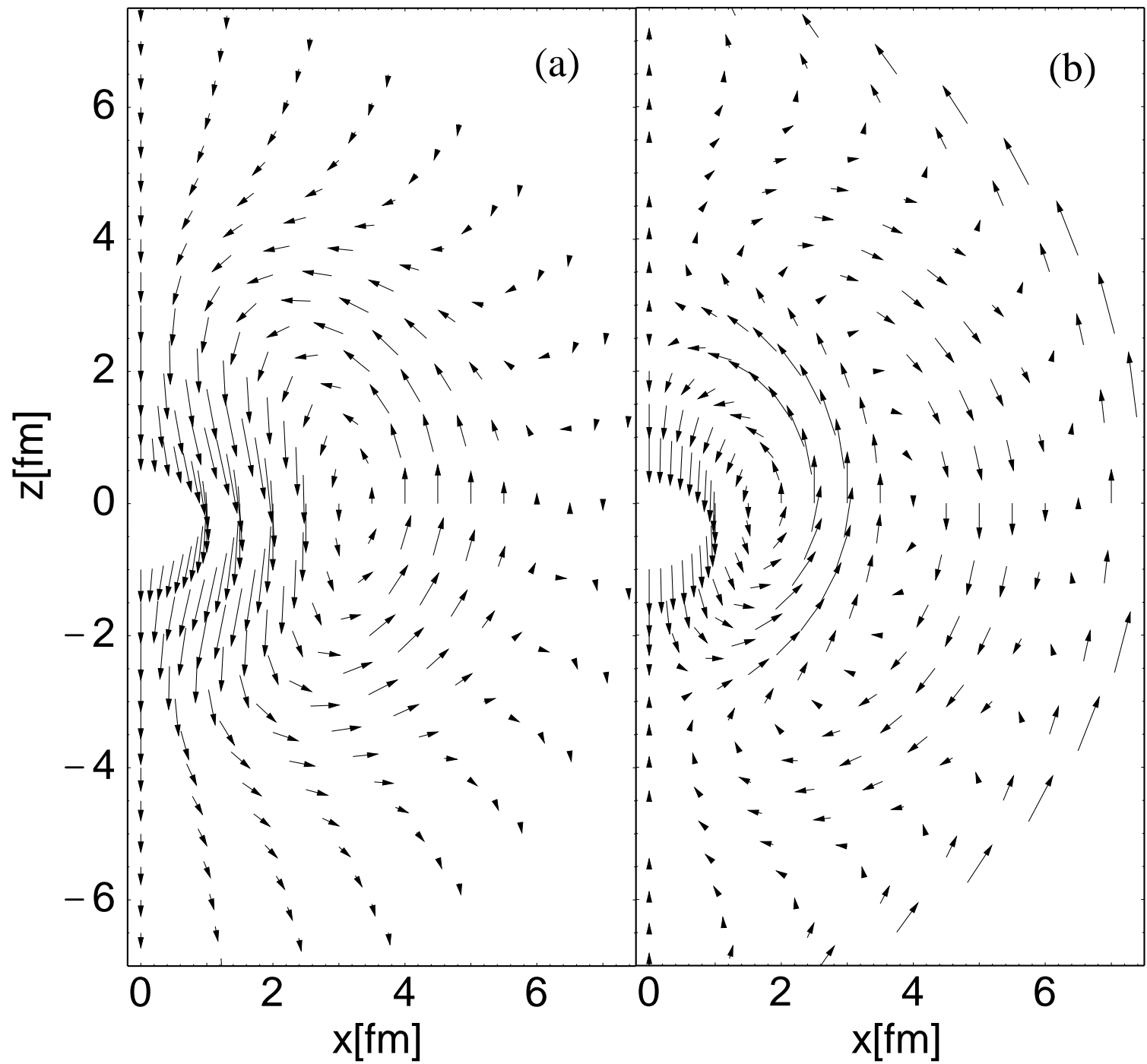
\end{document}